\documentclass[a4paper,onecolumn,oneside,11pt,leqno]{article}

\usepackage{amsmath, amsfonts,amssymb}
\usepackage{theorem, euscript,array,enumerate}
\usepackage{amsfonts,mathrsfs, latexsym}
\usepackage[dvips]{graphicx}
\usepackage[dvips]{graphics}
\usepackage[dvips]{epsfig}
\usepackage{color}

\theoremheaderfont{\bfseries} \theoremstyle{plain}
\theorembodyfont{\upshape}
\newcommand{\esp}{\mathbb{E}} 

\newcommand{\indic}{1\!\!1} 
\newcommand{\proba}{\mathbb{P}}

\newcommand{\R}{\mathbb{R}}
\newcommand{\D}{\mathcal{D}}
\newcommand{\F}{\mathcal{F}}
\newcommand{\G}{\mathcal{G}}

\def\lbr{[\![}

\DeclareMathOperator{\esssup}{esssup}
\theoremstyle{plain}

\theorembodyfont{\upshape\it}
\newtheorem{Thm}{\bf Theorem}[section]
\newtheorem{Pro}[Thm]{\bf Proposition}
\newtheorem{Lem}[Thm]{\bf Lemma}
\newtheorem{Cor}[Thm]{\bf Corollary}
\theorembodyfont{\upshape}

\newtheorem{Def}[Thm]{Definition}
\newtheorem{Exe}[Thm]{Example}
\newtheorem{Rem}[Thm]{Remark}
\newtheorem{Hyp}[Thm]{Assumption}

\newenvironment{proof}[1][{\it Proof.}]{\begin{trivlist}
\item[\hskip \labelsep {\bfseries #1}]}
{ {\quad}\hfill $\Box$\end{trivlist}\vskip
-0.2 cm}

\newcommand{\btau}{\boldsymbol{\tau}}
\newcommand{\bs}{\boldsymbol{s}}
\newcommand{\bL}{\boldsymbol{L}}

\begin{document}
\thispagestyle{empty}
\title{Multiple defaults and contagion risks }
\author{Ying Jiao\thanks{Laboratoire de probabilit\'es et mod\`eles
al\'eatoires, Universit\'e Paris 7,
jiao@math.jussieu.fr.
} }
\date{\today}
\maketitle
\begin{abstract}
We study multiple defaults where the global market information is
modelled as progressive enlargement of filtrations. We shall
provide a general pricing formula by establishing a relationship
between the enlarged filtration and the reference default-free
filtration in the random measure framework. On each default
scenario, the formula can be interpreted as a Radon-Nikodym
derivative of random measures. The contagion risks are studied in
the multi-defaults setting where we consider the optimal
investment problem in a contagion risk model and show that the
optimization can be effectuated in a recursive manner with respect
to the default-free filtration.
\end{abstract}

\section{Introduction}\label{Sec:introduction}

The contagion credit risk analysis with multiple default events is an important
issue for evaluating the credit derivatives and for the
 risk management facing the
financial crisis. Compared to the single credit name studies, there are
several difficulties in the multi-defaults context. Generally
speaking, the global market information containing all defaults
information is modelled as a recursive enlargement of filtrations
of all default times with respect to a default-free reference
filtration. To obtain the value process of a credit-sensitive
claim, one needs to consider the conditional expectation of its payoff function with respect to the global market
filtration. The mathematical formulation and computations are in
general complicated considering all possible default scenarios and the enlarged filtration.
Furthermore, the modelling of correlation structures of default
times in a dynamic manner is a challenging subject, in particular, when it concerns how to take into consideration the impact of one default event on the remaining names.

In the literature, there are mainly two approaches -- bottom up
and top down -- to model multiple default events. In the first
approach, one is interested in the probability distributions of
each individual default and in their correlations, often using
copula functions. The second approach concentrates directly on the
cumulative losses distributions, which allows to reduce the
complexity of the problem. However, the correlation structure
between the default times is not straightforward in the top-down
models. Recently, a new approach has been proposed to study the
successive default events (see \cite{ejj2}, also \cite{ES2009}),
which provides an intermediary point of view between the above two
approaches. The main ideas of \cite{ejj2} are two-folded. On one
hand, the default scenarios are largely reduced and we can
thoroughly analyze the impact of each default event on the
following ones; on the other hand, the computations are decomposed
on each default scenario and hence concern only the default-free
reference filtration. One key hypothesis is that the family of
default times admits a density with respect to the reference
filtration. The density hypothesis is a standard one in the
enlargement of filtrations (see for example \cite{jac85} for the
initial enlargement of filtration). In the credit risk analysis,
the density hypothesis has been adopted in \cite{ejj} for
analyzing what goes on after a default event, it has also been
proved to be useful in the recursive before-default and
after-default extensions.

Inspired by the second idea mentioned previously in \cite{ejj2},
we study the non-ordered multiple defaults by establishing a
formal relationship between the global market filtration and the
default-free reference filtration. We shall adopt  the framework
of random measures, which will provide us convenient and concise
notations. One can find a detailed introduction to random measures
in \cite[Chap II]{jac_shi} and in the monograph \cite{cra}.
Similar notions have also been used in the filtering problems (see
e.g. \cite{Kun} and \cite{Yor}). The main advantage of introducing
such a general framework is that we can treat the multiple
defaults case in a coherent  way as in the single name case.
Another consequence is that we can remove the density hypothesis.
In the case where explicit results are needed and where the
density hypothesis is assumed, we recover a result in
\cite{ejj2}, to which we refer for more detailed discussions.

As applications, we are interested in the pricing with multiple
defaults and in the contagion risks. The important idea in both
cases, as mentioned above, is to find a suitable decomposition of
the problem on each default scenario, so that the analysis will
only concern the default-free filtration. We shall present a
general pricing formula, which gives the value process with
respect to the global market information. On each default
scenario, the formula can be interpreted as a Radon-Nikodym
derivative of random measures. This result can be applied to
credit portfolio derivatives and also to contingent claims
subjected to contagion default risks.

The contagion credit risk during the financial crisis is an
important subject which needs to be taken into consideration.
Notably, one default event can have significant impact on
remaining firms on the market and may become  the potential cause
of other defaults, called the contagious defaults. We shall
present a contagion risk model to describe this phenomenon, where
each asset is influenced by the default risks of an underlying
portfolio and has a jump on its value at every default time. We
consider furthermore an investment portfolio containing such
assets and study the optimal investment strategy. We show that the
global optimization problem is equivalent to a family of recursive
optimization problems with respect to the default-free filtration.

The paper is organized as follows. We present in Section 2 the
mathematical framework of random measures. A general pricing
result concerning multiple defaults is deduced using random
measures in Section 3, and is applied to credit portfolio
derivatives. The Section 4 is devoted to the analysis on contagion
risks. We firstly present a multi-defaults contagion model, and
then study the optimal investment problems in the presence of
contagion risks.

\section{Random measure framework}
\subsection{Preliminaries}\label{sec prel}
Let $(\Omega,\mathcal G,\proba)$ be a probability space equipped
with a reference filtration $\mathbb F=(\F_t)_{t\geq 0}$
satisfying the usual conditions. Let
$\btau=(\tau_1,\cdots,\tau_n)$ be a family of random times taking
values in $\mathbb R_+^n$, considered as $\mathbb
R_+^{\{1,\cdots,n\}}$, representing the family of default times.
Denote by $\Theta$ the index set $\{1,\cdots,n\}$. We suppose that
$\tau_i$ $(i\in\Theta)$ are strictly positive and finite, and that
$\tau_i\neq \tau_j$, a.s. for $i\neq j$ $(j\in\Theta)$. Let
$\mathbb D^i=(\mathcal D_t^i)_{t\ge 0}$ be the smallest
right-continuous filtration such that $\tau_i$ is a $\mathbb
D^i$-stopping time. More precisely, $\mathcal
D_t^i:=\bigcap_{\varepsilon>0}\sigma(\tau_i\wedge
(t+\varepsilon))$. Let $\mathbb G=(\G_t)_{t\geq 0}$ be the
progressive enlargement of $\mathbb F$ by the default filtrations,
namely, $\mathbb G=\mathbb F\vee\mathbb D^1\vee\cdots\vee\mathbb
D^n$.

For any $I\subset\Theta$, let $\tau_I=(\tau_i)_{i\in I}$,
which is a random variable valued in $\mathbb R_+^I$. For
$t\in\mathbb R_+$, the notation $A^{I}_{t}$ denotes the event
\[A^{I}_{t}:=\bigg(\bigcap_{i\in I}\{\tau_i\le t\}\bigg)\cap
\bigg(\bigcap_{i\not\in I}\{\tau_i>t\}\bigg).\] The events
$(A^I_t)_{I\subset\Theta}$ describe all default scenarios
at time $t$.  Note that $\Omega$ is the disjoint union of $(A^I_t)_{I\subset\Theta}$.

The following lemma is an extension of a classical result on
progressive enlargement of filtrations with one default name.
\begin{Lem}\label{Lem:decomY}For $t\in\R_+$, any $\G_t$-measurable random variable $Y_t$ can be written in the decomposed form
\begin{equation}\label{decomp}
Y_t=\sum_{I\subset\Theta} \indic_{A_t^I} Y_t^I(\tau_I)
\end{equation} where $Y_t^I(\cdot)$ is a $\mathcal F_t\otimes\mathcal B(\mathbb R_+^I)$-measurable function on $\Omega\times\mathbb R_+^I$, $\mathcal B(\mathbb R_+^I)$ being the Borel $\sigma$-algebra.
\end{Lem}
\begin{proof}
By definition, for any $t\in\R_+$ and any integer $m>0$, the
random variable $Y_t$ is $\mathcal
F_t\vee\sigma(\boldsymbol{\tau}\wedge(t+1/m))$-measurable. Hence
there exists an $\mathcal F_t\otimes\mathcal B(\mathbb
R_+^n)$-measurable function $F_m$ such that
$Y_t(\omega)=F_m(\omega,\boldsymbol{\tau}\wedge(t+1/m))$. For
$I\subset\Theta$,
\[\indic_{A_t^I}Y_t=\indic_{\{\tau_I\le t,\,\tau_{I^c}>t\}}
F_m\Big(\omega,\btau\wedge(t+\frac 1m)\Big).\]
So, for fixed $\omega$, one has
\[\indic_{A_t^I}(\omega)Y_t(\omega)=\indic_{A_t^I}(\omega)F_m\Big(
\omega,\tau_I(\omega), \big(t+\frac 1m\big)_{I^c}\Big)\] when $m$
is large enough. Let
\[Y_t^I(s_I):=\displaystyle\limsup_{m\rightarrow\infty}
F_m(\omega,\boldsymbol{x}^{(m)})\] where
$\boldsymbol{x}^{(m)}=(x_1^{(m)},\cdots,x_n^{(m)})$ is defined as
$x_i^{(m)}:= s_i$ if $i\in I$ and $x_i^{(m)}:= t+1/m$ if $i\not\in
I$. Then one has $\indic_{A_t^I}Y_t=\indic_{A_t^I}Y_t^I(\tau_I)$.
\end{proof}

The following variant of Lemma \ref{Lem:decomY} will be useful further on.
\begin{Lem}\label{Lem:decomY2}
{Any $\mathcal G_{t-}$-measurable random variable $Y_t$ can be
written as
\[Y_t=\sum_{I\subset\Theta}\indic_{A^I_{t-}}Y_t^I(\tau_I),\]
where \[A^I_{t-}:=\bigg(\bigcap_{i\in I}\{\tau_i< t\}\bigg)\cap\bigg(\bigcap_{i\not\in I}\{\tau_i\ge t\}\bigg),\]
and $Y_t^I(\cdot)$ is $\mathcal F_{t-}\otimes\mathcal B(\mathbb R_+^I)$-measurable.}
\end{Lem}
\begin{proof}
For any $t>0$, $\mathcal G_{t-}=\bigcup_{\varepsilon>0} \mathcal
F_{t-\varepsilon}\vee\sigma(\btau\wedge(t-\varepsilon))$. So there
exists an integer $m>0$ such that $Y_t$ is $\mathcal
F_{t-}\otimes\sigma(\btau\wedge(t-1/m))$-measurable. Hence $Y_t$
can be written as $F_m(\omega,\btau\wedge(t-1/m))$ where $F_m$ is
an $\mathcal F_{t-}\otimes\mathcal B(\mathbb R_+^n)$-measurable
function. Then we can complete the proof by a similar argument as
for Lemma \ref{Lem:decomY}.
\end{proof}

\begin{Rem}
In the Lemmas \ref{Lem:decomY} and \ref{Lem:decomY2}, if the
random variable $Y_t$ is positive (resp. bounded), then
$Y_t^I(\cdot)$ can be chosen to be positive (resp. bounded).
\end{Rem}

\subsection{Random measures}
\begin{Def}
Let $\mu^{\btau}$ be the measure on $(\Omega\times\mathbb
R_+^n,\mathcal F_\infty\otimes\mathcal B(\mathbb R_+^n))$ such that for any positive and $\F_\infty\otimes\mathcal
B(\R_+^n)$-measurable function $h_\infty(\cdot)$,\begin{equation}\label{mu btau}\int h_\infty(\boldsymbol{s})\mu^{\btau}(d\omega,d\boldsymbol{s})
=\mathbb E[h_\infty(\btau)],\end{equation} where $\mathcal F_\infty=\bigcup_{t\ge 0}\mathcal F_t$ and $\bs=(s_1,\cdots,s_n)$.
\end{Def}

The measure $\mu^{\btau}$ can be considered as a transition kernel from $(\Omega,\mathcal F_{\infty})$ to $(\mathbb R_+^n,\mathcal B(\mathbb R_+^n))$ whose marginal on $\Omega$ coincides with $\mathbb P$. It can also be considered as the conditional law of $\btau$ on $\mathcal F_\infty$.
We give below an example  of $\mu^{\btau}$  using the copula model in \cite{SS}.
\begin{Exe}
(Sch\"{o}nbucher and Schubert) For any $i\in\Theta$, define the default time by $\tau_i=\inf\{t: \Lambda_t^i\geq U_i\}$ where $\Lambda^i$ is an continuous increasing   $\mathbb F$-adapted process and $U_i$ is an exponential distributed random variable independent of $\F_\infty$. The conditional survival probability is $u_t^i=\proba(\tau_i>t|\F_\infty)=\exp(-\Lambda_t^i)$. Note that H-hypothesis holds in this model, that is, $\proba(\tau_i>t|\F_t)=\proba(\tau_i>t|\F_\infty)$. The construction of joint survival distribution in \cite{SS} is by introducing a copula function $C: \R_+^n\rightarrow \R_+$ such that
$$\proba(\btau>\bs|\F_\infty)=\proba(\tau_1>s_1,\cdots,\tau_n>s_n|\F_\infty)=C(u^1_{s_1},\cdots,u^n_{s_n}).$$
Then for any positive and $\F_\infty\otimes\mathcal
B(\R_+^n)$-measurable function $h_\infty(.)$, one has
\[\mathbb E[h_\infty(\btau)]=\int h_\infty(\bs)\mu^{\btau}(d\omega,ds)=\esp\Big[\int_{\R_+^n}h_\infty(\bs)(-1)^nd_{s_1}\cdots d_{s_n}C(u_{s_1}^1,\cdots,u_{s_n}^n)\Big]
\]where $d_{s_1}\cdots d_{s_n}C(u_{s_1}^1,\cdots,u_{s_n}^n)$ is an $n$-dimensional Lebesgue-Stieltjes measure associated to $C(u_{s_1}^1,\cdots,u_{s_n}^n)$.
\end{Exe}

Classically the random measure is a straightforward extension of the notions of increasing processes and their compensators, see \cite[Chap II]{jac_shi} for details. Here the random measure $\mu^{\btau}$ is useful to define auxiliary measures on suitable $\sigma$-fields.
For $t\ge 0$, let $\mu_t^{\btau}$ be the restriction of $\mu^{\btau}$ on $\mathcal F_t\otimes\mathcal B(\mathbb R_+^n)$. It represents the conditional law of $\btau$ on $\mathcal F_t$. For this reason, we also write $\mu^{\btau}_t$ as $\mathbb E[\mu^{\btau}|\mathcal F_t]$. If $h_t(\bs)$ is a positive $\mathcal F_t\otimes\mathcal B(\mathbb R_+^n)$-measurable function, then
\[\mathbb E[h_t(\btau)]=\int h_t(\bs)
\mu^{\btau}(d\omega,d\bs)=\int h_t(\bs)\mu_t^{\btau}(d\omega,d\bs).\]
For $I\subset\{1,\cdots,n\}$, let $\mu^I_t$ be the measure on
$(\Omega\times\mathbb R_+^I,\mathcal F_t \otimes\mathcal B(\mathbb
R_+^I))$ which is a partial marginal measure of $\mu^{\btau}_t$ such
that for any positive and $\F_t\otimes\mathcal
B(\R_+^I)$-measurable function $h_t(s_I)$, $s_I=(s_i)_{i\in
I}$, one has
\begin{equation}\label{Equ:mu_I}
\int_{\Omega\times\R_+^I} h_t(
s_I)\mu^I_{t}(d\omega,ds_I)=\int_\Omega\int_{\R_+^I\times
]t,\infty[^{I^c}}h_t(s_I)
\mu^{\btau}_t(d\omega,d\boldsymbol{s}).\end{equation}
This relation can also be written as
\begin{equation}\label{Equ:muti}\mu_t^I(d\omega,ds_I)=\int_{]t,\infty[^{I^c}}\mu^{\btau}_t(d\omega,
d\bs).\end{equation}
For  $T\ge
t$, $I\subset\{1,\cdots,n\}$ and $Y_T(\cdot)$ which
is positive and $\mathcal F_T\otimes\mathcal B(\mathbb
R_+^n)$-measurable, we define $\mu_t^{Y_T,I}$ as the weighted
marginal measure on $(\Omega\times\mathbb R_+^I,\mathcal
F_t\otimes\mathcal B(\mathbb R_+^I))$ such that
\begin{equation}\label{Equ:mu_YI}\int_{\Omega\times \R_+^I} h_t(s_I)\mu^{Y_T,I}_t(d\omega,ds_I)=
\int_\Omega\int_{\R_+^I\times]t,\infty[^{I^c}}h_t(s_I)Y_T(\boldsymbol{s}) \mu^{\btau}(d\omega,d\bs).\end{equation}
Similarly, we write $\mu^{Y_T,I}_t$ as
\begin{equation}\label{Equ:mutii}\mu^{Y_T,I}_t(d\omega,ds_I)=\int_{]t,\infty[^{I^c}}\mathbb E[Y_T(\bs)\mu^{\btau}|\mathcal F_t](d\omega,d\bs),\end{equation}
where $E[Y_T(\bs)\mu^{\btau}|\mathcal F_t]$ denotes the restriction of the measure $Y_T(\bs)\mu^{\btau}$ on $\mathcal F_t\otimes\mathcal B(\mathbb R_+^n)$. Note that one has $E[Y_T(\bs)\mu^{\btau}|\mathcal F_t]=E[Y_T(\bs)\mu_T^{\btau}|\mathcal F_t]$.

We shall use the Radon-Nikodym derivative of random measures to interpret diverse conditional expectations.

\begin{Pro}\label{Pro:EspF}
Let $T\ge t\ge 0$. For any positive and $\mathcal
F_T\otimes\mathcal B(\mathbb R_+^n)$-measurable function
$Y_T(\cdot)$ on $\Omega\times\mathbb R_+^n$, one has
\begin{equation}\label{Equ:espf}\mathbb E[Y_T(\btau)|\mathcal
F_t]=\frac{\int_{\mathbb R_+^n} \mathbb
E[Y_T(\bs)\mu^{\btau}|\mathcal F_t] (d\omega,d\bs)}{\int_{\mathbb
R_+^n}\mu^{\btau}(d\omega,d\bs)}\end{equation} where the
Radon-Nikodym derivative is taken on $\F_t$.
\end{Pro}
\begin{Rem}
Note that $\int_{\mathbb R_+^n}\mu^{\btau}(d\omega,d\bs)=\mathbb
P(d\omega)$, the above result can be written as
\[\mathbb E[Y_T(\btau)|\mathcal F_t]\mathbb P(d\omega)=\int_{\mathbb R_+^n}
\mathbb E[Y_T(s_I)\mu_T^{\btau}|\mathcal F_t] (d\omega,d\bs).\]
\end{Rem}
\begin{proof}
Let $h_t$ be a positive $\mathcal F_t$-measurable random variable, then
\[\int h_t(\omega)\mathbb E[Y_T(\btau)|\mathcal F_t]\mathbb P(d\omega)=
\mathbb E\big[h_t\mathbb E[Y_T(\btau)|\mathcal F_t]\big] =\mathbb
E[h_tY_T(\btau)]= \int h_t(\omega)Y_T(\bs)\mu^{\btau}(
d\omega,d\bs).\] Hence the equality \eqref{Equ:espf} holds.
\end{proof}

\begin{Rem}\label{Rem2.8}
In particular, the conditional expectation $\mathbb E[Y_T|\mathcal F_t]$ where $Y_T$ is a positive $\mathcal G_T$-measurable random variable can be written in a decomposed form. In fact, by lemma \ref{Lem:decomY}, one has
\[Y_T=\sum_{I\subset\Theta}\indic_{A_T^I}Y_T^I(\tau_I)
=\sum_{I\subset\Theta}\indic_{[0,T]^I\times]T,\infty[^{I^c}}(
\btau)Y_T^I(\tau_I),\]
where $Y_T^I(\cdot)$ is positive and $\mathcal F_T\otimes\mathcal B(\mathbb R_+^I)$-measurable.
Hence Proposition \ref{Pro:EspF} gives
\[\mathbb E[Y_T(\btau)|\mathcal
F_t]\mathbb
P(d\omega)=\sum_{I\subset\Theta}\int_{[0,T]^I\times]T,\infty[^{I^c}}
\mathbb E[Y_T(s_I)\mu_T^{\btau}|\mathcal F_t] (d\omega,d\bs).\]
\end{Rem}

\section{Pricing with multiple defaults}

For the purpose of pricing, let us consider a contingent claim
sensitive to multiple defaults with the payoff function
$Y_T(\btau)$ where $Y_T(\bs)$ is a positive and
$\F_T\otimes\mathcal B(\R_+^n)$-measurable function on
$\Omega\times\R_+^n$, $T$ being the maturity. Since
$\btau=(\tau_1,\cdots,\tau_n)$ represents  a family of default
times, $Y_T(\btau)$ can describe a large class of financial
products such as a basket credit derivative, or a single-name
contingent claim subjected to the default risks of multiple
counterparties, or a basket European option with contagion risks
etc. The price of this product is computed as the expectation
$\esp[Y_T(\btau)]$ under some risk-neutral probability measure.
The dynamic price process given all market information  at time
$t\leq T$ is the conditional expectation  $\esp[Y_T(\btau)|\G_t]$.
In this section, we shall present the evaluation formulas using
the random measures.

\subsection{General pricing formula}

We suppose in this section that the (conditional) expectations are taken under some risk-neutral probability.

\begin{Thm}\label{Thm:main thm} Let $T\ge t\ge 0$.
For any positive and $\mathcal F_T\otimes\mathcal B(\mathbb
R_+^n)$-measurable function $Y_T(\cdot)$ on $\Omega\times\R_+^n$,
the measure $\mu^{Y_T,I}_t$ is absolutely continuous with respect
to $\mu^I_t$. Moreover, the following equality holds
\begin{equation}\label{Equ:main}
\mathbb E[Y_T(\btau)|\mathcal G_t]=
\sum_{I\subset\Theta}\indic_{A_t^I}\frac{d\mu^{Y_T,I}_t}{d\mu^I_t}(\omega,\tau_I).
\end{equation}
Using the notations \eqref{Equ:muti} and \eqref{Equ:mutii}, the
above equality can also  be written as
\begin{equation}\label{Equ:main2}\mathbb E[Y_T(\btau)|\mathcal G_t]=\sum_{I\subset\Theta}\indic_{A_t^I}\frac{\int_{]t,\infty[^{I^c}}
\mathbb E[Y_T^I(s_I)\mu^{\btau}_T|\mathcal F_t](d\omega,d\bs)}{
\int_{]t,\infty[^{I^c}}\mu^{\btau}_t(d\omega,d\bs)}\bigg|_{s_I=\tau_I}
\end{equation}
where the Radon-Nikodym derivative is taken on the $\sigma$-algebra $\mathcal F_t\otimes\mathcal B(\mathbb R_+^I)$.
\end{Thm}
\begin{proof}
By definition \eqref{Equ:mu_I}, one has that, for any $M\in\mathcal F_t\otimes\mathcal B(\mathbb R_+^I)$, $\mu_t^I(M)=0$ if and only if $\indic_M(\omega,s_I)\indic_{
]t,+\infty[^{I^c}}(s_{I^c})=0$, $\mu^{\btau}$-a.e. Hence this implies $\mu_t^{Y_T,I}(M)=0$. \\
On the set $A_t^I$, any $\G_t$-measurable test random variable can be written in the form $Z_t(\tau_I)$, where $Z_t(s_I)$ is positive  and $\F_t\otimes\mathcal B(R_+^I)$-measurable. To prove \eqref{Equ:main}, it suffices to establish
\begin{equation}\label{Equ:main1}\mathbb E[\indic_{A_t^I}Z_t(\tau_I)Y_T(\tau)]=
\mathbb E\Big[\indic_{A_t^I}Z_t(\tau_I)
\frac{d\mu_t^{Y_T,I}}{d\mu_t^I}(\omega,\tau_I)\Big].\end{equation}
One has
\[\begin{split}
&\quad\;\mathbb E\Big[\indic_{A_t^I}Z_t(\tau_I)
\frac{d\mu_t^{Y_T,I}}{d\mu_t^I}(\omega,\tau_I)\Big]\\&=
\int\indic_{[0,t]^I}(s_I)\indic_{]t,+\infty[^{I^c}}(s_{I^c})
Z_t(s_I)\frac{d\mu_t^{Y_T,I}}{d\mu_t^I}(\omega,s_I)\mu^{\btau}(d\omega,
d\bs)\\
&=\int \indic_{[0,t]^I}(s_I)Z_t(s_I)\frac{d\mu_t^{Y_T,I}}{d\mu_t^I}(\omega,s_I)
\mu_t^I(d\omega,ds_I)\\
&=\int\indic_{[0,t]^I}(s_I)Z_t(s_I)\mu_t^{Y_T,I}(d\omega,ds_I)\\
&=\int\indic_{[0,t]^I}(s_I)\indic_{]t,+\infty[^{I^c}}(s_{I^c})
Z_t(s_I)Y_T(\bs)\mu^{\btau}(d\omega,d\bs),
\end{split}\]
where the first equality comes from the definition of $\mu^{\btau}$, the second one comes from \eqref{Equ:mu_I}, the third one results from the definition of Radon-Nikodym derivative, and the last one comes from \eqref{Equ:mu_YI}. Again by the definition of $\mu^{\btau}$, the last formula equals the left side of \eqref{Equ:main1}.
\end{proof}

\begin{Rem}
The following form of Theorem \ref{Thm:main thm} will be useful. Let $Y_T$ be a positive $\mathcal G_T$-measurable random variable, which is written as
\[Y_T=\sum_{I\subset\Theta}\indic_{A_T^I}Y_T^I(\tau_I),\]
where $Y_T^I(\cdot)$ is $\mathcal F_T\otimes\mathcal B(\mathbb R_+^I)$-measurable. Theorem \ref{Thm:main thm} implies that
\begin{equation}\begin{split}\mathbb E[Y_T|\mathcal G_t]&=\sum_{I\subset\Theta}\esp[\indic_{A_T^I}Y_T^I(\tau_I)|\G_t]\\&=
\sum_{I\subset\Theta}\sum_{J\subset\Theta}\indic_{A_t^J}
\frac{\int_{]t,\infty[^{J^c}}\mathbb E[\indic_{[0,T]^I
\times]T,\infty[^{I^c}}(\bs)Y_T^I(s_I)\mu_T^{\btau}|\mathcal F_t]
(d\omega,d\bs)}{\int_{]t,\infty[^{J^c}}\mu_t^{\btau}(d\omega,d\bs)}\\
&=\sum_{I\subset\Theta}\sum_{J\subset I}
\indic_{A_t^J}\frac{\int_{]t,\infty[^{J^c}}\indic_{[0,T]^I
\times]T,\infty[^{I^c}}(\bs)\mathbb E[Y_T^I(s_I)\mu_T^{\btau}|\mathcal F_t]
(d\omega,d\bs)}{\int_{]t,\infty[^{J^c}}\mu_t^{\btau}(d\omega,d\bs)}\\
&=\sum_{J\subset\Theta}\indic_{A_t^J}
\sum_{I\supset J}\frac{\int_{]T,\infty[^{I^c}\times
]t,T]^{I\setminus J}}\mathbb E[Y_T^I(s_I)\mu_T^{\btau}|\mathcal F_t](d\omega,d\bs)}{\int_{]t,\infty[^{J^c}}\mu_t^{\btau}(d\omega,d\bs)},
\end{split}\end{equation}
where the last equality comes from an interchange of summation.
\end{Rem}

Inspired by \cite{ejj2}, we consider the case where the density hypothesis holds. Let
$\nu^{\btau}$ be the marginal measure of $\mu^{\btau}$ on
$\mathcal B(\R_+^n)$, that is,
$$\nu^{\btau}(U)=\mu^{\btau}(\Omega\times U), \quad \forall\, U\in\mathcal B(\R_+^n).$$
Note that $\nu^{\btau}$ is actually the law of $\btau$.
\begin{Hyp} \label{hyp:density}We say that $\btau=(\tau_1,\cdots,\tau_n)$ satisfies the {\it density hypothesis} if the measure $\mu^{\btau}$  is absolutely continuous with respect to  $\proba\otimes \nu^{\btau}$. We denote by $\alpha_t(\cdot)$ the density of $\mu^{\btau}$ with respect to $\mathbb P\otimes\nu^{\btau}$ on $(\Omega\times\R_+^n, \F_t\otimes\mathcal B(\R_+^n))$, where $t\in\mathbb R_+$.
\end{Hyp}
Under the above density hypothesis, one has, for any positive Borel
function on $\R_+^n$,
$$\esp[f(\btau)|\F_t]=\int_{\R_+^n}f(\bs)\alpha_t(\bs)\nu^{\tau}(ds).$$
This relationship can also be written as
\begin{equation}\label{Equ:dens}\mu^{\btau}_t(d\omega,d\bs)=\alpha_t(\bs)\mathbb P(d\omega)
\otimes\nu^{\btau}(d\bs).\end{equation}

\begin{Cor}We keep the notation of Theorem \ref{Thm:main thm} and assume in addition the density hypothesis \ref{hyp:density}.
Then \begin{equation}\label{prix density} \mathbb
E[Y_T(\btau)|\mathcal G_t]=
\sum_{I}\indic_{A_t^I}\frac{\int_{]t,+\infty[^{I^c}}\esp[Y_T(\bs)\alpha_T(\bs)|\F_t]\nu^{\btau}(d\bs)}{\int_{]t,\infty[^{I^c}}\alpha_t(\bs)\nu^{\btau}(d\bs)}\Big|_{s_I=\tau_I}
\end{equation}
\end{Cor}
\begin{proof}
By the density hypothesis \eqref{Equ:dens} and by \eqref{Equ:muti}, for any $I\subset\Theta$,  one has
 \[\mu_t^I(d\omega,ds_I)=\int_{]t,\infty[^{I^c}}\mu_t^{\btau}(d\omega,d\bs)=
\int_{]t,\infty[^{I^c}}\alpha_t(\bs)\mathbb P(d\omega)
\otimes\nu^{\btau}(d\bs).
\]
Similarly,
\[\mu_t^{Y_T,I}(d\omega,ds_I)=\int_{]t,\infty[^{I^c}}\mathbb E[Y_T(s_I)\mu_T^{\btau}
|\mathcal F_t](d\omega,d\bs)=\int_{]t,\infty[^{I^c}}
\mathbb E[Y_T(s_I)\alpha_T(\bs)|\mathcal F_t]\mathbb P(d\omega)\otimes\nu^{\btau}(d\bs).\]
So
\[\frac{d\mu_t^{Y_T,I}}{d\mu_t^I}=
\frac{\int_{]t,\infty[^{I^c}}\esp[Y_T(\bs)\alpha_T(\bs)|\F_t]
\nu^{\btau}(d\bs)}{\int_{]t,\infty[^{I^c}}\alpha_t(\bs)
\nu^{\btau}(d\bs)}.\]
In fact, for any positive $\mathcal F_t\otimes\mathcal B(\mathbb R_+^I)$-measurable function $h_t(\cdot)$, one has
\[\begin{split}
&\quad\;\int_{\Omega\times\mathbb R_+^I}h_t(s_I)\frac{\int_{]t,+\infty[^{I^c}}\esp[Y_T(\bs)\alpha_T(\bs)|\F_t]\nu^{\btau}(d\bs)}{\int_{]t,\infty[^{I^c}}\alpha_t(\bs)\nu^{\btau}(d\bs)}
\mu_t^I(d\omega,ds_I)\\
&=\int_{\Omega\times\mathbb R_+^I\times]t,\infty[^{I^c}}h_t(s_I)\frac{\int_{]t,+\infty[^{I^c}}\esp[Y_T(\bs)\alpha_T(\bs)|\F_t]\nu^{\btau}(d\bs)}{\int_{]t,\infty[^{I^c}}\alpha_t(\bs)\nu^{\btau}(d\bs)}
\alpha_t(\bs)\mathbb P(d\omega)
\otimes\nu^{\btau}(d\bs)\\
&=\int_{\Omega}\mathbb P(d\omega)\int_{\mathbb R_+^I}h_t(s_I)\frac{\int_{]t,+\infty[^{I^c}}\esp[Y_T(\bs)\alpha_T(\bs)|\F_t]\nu^{\btau}(d\bs)}{\int_{]t,\infty[^{I^c}}\alpha_t(\bs)\nu^{\btau}(d\bs)}
\int_{]t,\infty[^{I^c}}\alpha_t(\bs)\nu^{\btau}(d\bs)\\
&=\int_{\Omega}\mathbb P(d\omega)\int_{\mathbb R^I\times]t,\infty[^{I^c}}h_t(s_I)\mathbb E[Y_T(\bs)\alpha_T(\bs)|\mathcal F_t]\nu^{\btau}(d\bs)\\
&=\int_{\Omega\times\mathbb R_+^I\times]t,\infty[^{I^c}}
h_t(s_I)Y_T(\bs)\alpha_T(\bs)\mathbb P(d\omega)\otimes\nu^{\btau}(d\bs)\\
&=\int_{\Omega\times\mathbb R_+^I}h_t(s_I)\mu_t^{Y_T,I}(d\omega,ds_I),
\end{split}\]
where the first and the last equality come from the density
hypothesis, the second and the fourth ones come from Fubini's
theorem. Thus the equality \eqref{prix density} follows from
Theorem \ref{Thm:main thm}.
\end{proof}

\subsection{Pricing of credit portfolio derivatives}
We now apply the previous pricing formulas to the two important types of credit portfolio derivatives: the
$k^{\text{th}}$-to-default swaps and the CDOs.

\subsubsection{Basket default swaps}
A $k^{\text{th}}$-to-default swap provides to its buyer the
protection against the $k^{\text{th}}$ default of the underlying
portfolio. Let $(\tau_{(k)})_{k\in\Theta}$ be the ordered set of
$\btau=(\tau_i)_{i\in\Theta}$, that is,
$\tau_{(1)}<\cdots<\tau_{(n)}$. The protection buyer pays a
regular premium until the occurrence of the $k^{\text{th}}$
default time $\tau_{(k)}$ or until the maturity $T$ if there are
less than $k$ defaults before $T$. In return, the protection
seller pays the loss $1-R_{(k)}$ where $R_{(k)}$ is the recovery
rate if $\tau_{(k)}\leq T$, and pays zero otherwise. So the key
term  for evaluating such a product is the indicator default
process $\indic_{\{\tau_{(k)}\leq T\}}$ with respect to the market
filtration $\G_t$.

\begin{Pro}\label{Pro:kdef}
 For any $t\leq T$,
\begin{equation}\label{Equ:kth}
\esp[\indic_{\{\tau_{(k)}>T\}}|\G_t]=\sum_{|J|< k}
\indic_{A_t^J}\sum_{I\supset J,\,|I|< k}\frac{\int_{]T,\infty[^{I^c}}\int_{]t,T]^{I\setminus J}}\mu_t^{\boldsymbol{\tau}}(d\omega,d\boldsymbol{s})}{
\int_{]t,\infty[^{J^c}}\mu^{\btau}_t(d\omega,d\boldsymbol{s})}\bigg|_{s_J=\tau_J}.\end{equation}
\end{Pro}
\begin{proof}
Observe that $\indic_{\{\tau_{(k)}>T\}}=\sum_{|I|<
k}\indic_{A_T^I}$. By Theorem \ref{Thm:main thm}, one obtains
\[
\mathbb E[\indic_{A_T^I}|\mathcal G_t]=\sum_{J\subset I}\indic_{A_t^J}\frac{\int_{]T,\infty[^{I^c}}
\int_{]t,T]^{I\setminus J}}\mu_t^{\boldsymbol{\tau}}
(d\omega,d\boldsymbol{s})}{\int_{]t,\infty[^{J^c}}
\mu_t^{\boldsymbol{\tau}}(d\omega,d\boldsymbol{s})}\bigg|_{s_J=\tau_J}.\]
By taking the sum over $I$ such that $|I|\le k$ and by interchanging the summations, one gets \eqref{Equ:kth}.
\end{proof}

\begin{Rem}
Among the basket default swaps, the \emph{first-to-default swap} is the most important one.  In this case, $k=1$. Proposition \ref{Pro:kdef} leads to
\[\mathbb E[\indic_{\{\tau_{(1)}>T\}}|\mathcal G_t]
=\indic_{\{\tau_{(1)}>t\}}\frac{\int_{]T,\infty[^{n}}\mu_t^{\boldsymbol{\tau}}(d\omega,d\boldsymbol{s})}{
\int_{]t,\infty[^{n}}\mu^{\btau}_t(d\omega,d\boldsymbol{s})}=\indic_{\{\tau_{(1)}>t\}}\frac{\mathbb E[\indic_{\{\tau_{(1)}>T\}}|\mathcal F_t]}{\mathbb E[\indic_{\{\tau_{(1)}>t\}}|\mathcal F_t]}.\]The last equality is a well-known result (e.g. \cite{BJR}).
\end{Rem}

\subsubsection{CDO tranches}
A CDO is a structured credit derivative based on a large pool of
underlying assets and containing several tranches.  For the
pricing of a CDO tranche, the term of interest is the cumulative
loss of the portfolio $l_t=\sum_{i=1}^nR_i\indic_{\tau_i\leq t}$,
$R_i$ being the recovery rate of $\tau_i$. A tranche of CDO is
specified by an interval corresponding to the partial loss of the
portfolio. The two threshold values, a upper value $a$ and a lower
one $b$, defines a tranche of CDO and the loss on the tranche is
given as a call spread written on the loss process $l_t$ with
strike values $a$ and $b$. Therefore to obtain the dynamics of the
CDO prices, we shall consider $\esp[(l_T-a)^+|\G_t]$.

On the market, it is a standard hypothesis to suppose that the
recovery rate for each underlying name is constant (equal to
$40\%$ in practice). We make this hypothesis below and  discuss
further on in Remark \ref{Rem fin}  the case where the recovery
rates $R_i$ are random variables.

The following result allows us to deduce the CDO prices using the
$k^{\text{th}}$-to default swaps.
\begin{Pro}
Assume that $R_i=R$ is constant for all $i\in\Theta$, then
\begin{equation}\label{CDO}(l_T-a)_+=R\sum_{k\ge a/R}\mathbb
\min(k-\frac{a}{R},1)\,\indic_{\{\tau_{(k)}\leq
T\}}.\end{equation} Moreover,
\[\mathbb E[(l_T-a)_+|\mathcal G_t]=R\sum_{k\ge a/R}\mathbb \min(k-\frac{a}{R},1)\bigg[1-\sum_{|J|< k}
\indic_{A_t^J}\sum_{I\supset J,\,|I|<
k}\frac{\int_{]T,\infty[^{I^c}}\int_{]t,T]^{I\setminus
J}}\mu^{\boldsymbol{\tau}}(d\omega,d\boldsymbol{s})}{
\int_{]t,\infty[^{J^c}}\mu^{\btau}(d\omega,d\boldsymbol{s})}\bigg|_{s_J=\tau_J}\bigg].\]
\end{Pro}
\begin{proof}Notice  that for any $m\in\Theta$, the
following equality holds
\[(m-\frac aR)_+=\sum_{\begin{subarray}{c}
k\in\Theta, \frac aR\leq k\leq m\end{subarray}}\min(k-\frac aR,
1).\]Hence
\[(\sum_{i=1}^n\indic_{\{\tau_i\leq T\}}-\frac aR)_+=
\sum_{k\in\Theta, \frac aR\leq k}\min(k-\frac aR,
1)\indic_{\{k\leq \sum_{i=1}^n\indic_{\{\tau_i\leq T\}}\}}.\]
Since $\{k\leq \sum_{i=1}^n\indic_{\{\tau_i\leq
T\}}\}=\{\tau_{(k)}\leq T\}$, we obtain \eqref{CDO} and
\[\mathbb E[(l_T-a)_+|\mathcal G_t]=\sum_{k\ge \frac aR}\mathbb \min(k-\frac aR,1)\,\esp[\indic_{\{\tau_{(k)}\le T\}}|\mathcal G_t].\]
By Proposition \ref{Pro:kdef}, the result holds.
\end{proof}

\section{The contagion risk model}
In this section, we are interested in the contagion risks with
multiple defaults. One observation during the financial crisis is
that one default event may have impact on other remaining firms
and often causes important losses on the asset values of its
counterparties. We shall propose a contagion model to take into
consideration this phenomenon.

\subsection{Preliminaries}
We begin by generalizing  Lemmas \ref{Lem:decomY} and
\ref{Lem:decomY2} to the case of processes. Denote by $\mathcal
O_{\mathbb F}$ (resp. $\mathcal P_{\mathbb F}$) the optional
(resp. predictable) $\sigma$-field on $\Omega\times\mathbb R_+$.

\begin{Lem}\label{Lem:decomopt} 1) Any $\mathbb G$-optional process $Y$ can be written as
$Y_t=\sum_I \indic_{A_t^I} Y_t^I(\tau_I)$
where $Y^I(\cdot)$ is an $\mathcal O_{\mathbb F}\otimes\mathcal B(\mathbb R_+^I)$-measurable function on $\Omega\times\mathbb R_+\times\mathbb R_+^I$.\\
2) Any $\mathbb G$-predictable process $Y$ can be written as
$Y_t=\sum_I \indic_{A_{t-}^I} Y_t^I(\tau_I)$ where $Y^I(\cdot)$ is a $\mathcal P_{\mathbb F}\otimes\mathcal B(\mathbb R_+^I)$-measurable function on $\Omega\times\mathbb R_+\times\mathbb R_+^I$.
\end{Lem}
\begin{proof}
1) It suffices to consider $Y=Z\indic_{\lbr s,\infty\lbr}$, $Z$ being a $\mathcal G_s$-measurable random variable. By Lemma \ref{Lem:decomY}, for any $I\subset\Theta$, there exists an $\mathcal F_s\otimes\mathcal B(\mathbb R_+^I)$-measurable function $Z^I(\cdot)$ such that
\[Z=\sum_{I\subset\Theta}\indic_{A_s^I}Z^I(\tau_I).\]
We define
\[Y_t^I(s_I):=
\begin{cases}\sum_{J\subset I} Z^J(s_J)
\indic_{[0,s]^J}(s_J)\prod_{i\in I\setminus J}1_{\{s<s_i\le t\}}\quad &\text{if }t\ge s,\\
0  &\text{if } t<s. \end{cases}\] Notice that the process
$Y^I(s_I)$ is right continuous for any $s_I$. Hence $Y^I$ is
$\mathcal O_{\mathbb F}\otimes \mathcal B(\mathbb
R_+^I)$-measurable. By the equality
\[\sum_{I\supset J}\indic_{A_t^I}\Big(\prod_{i\in I\setminus J}
\indic_{\{s< \tau_i\le
t\}}\Big)\indic_{[0,s]^J}(\tau_J)=\indic_{A_s^J}\] which holds for
any $J\subset\Theta$ and any $t\in[s,+\infty[$, one can verify
that
$Y_t=\sum_I\indic_{A_t^I}Y^I(\tau_I)$.\\
2) By using  Lemma \ref{Lem:decomY2}, a variant of the above
argument leads to the predictable version of 1).

\end{proof}

\subsection{The model setup}
We consider a portfolio of $N$ assets, whose value process is  denoted by $S$ which is an $N$-dimensional $\mathbb G$-adapted process. The process $S$ has the following decomposed form
\[S_t=\sum_{I\subset\Theta}\indic_{A_t^I}S_t^{I}(\tau_I),\]
where $S_t^{I}(s_I)$ is $\mathcal O_{\mathbb F}\otimes \mathcal
B(\R_+^I)$-measurable and takes values in $\R_+^N$, representing
the asset values given the past default events $\tau_I=s_I$. Note
that $S_t$ depends on the value of $S_t^I(\cdot)$ only on the set
$A_t^I$, that is, only when $t\geq\max s_I$. Hence we may assume
in convention that $S_t^I(s_I)=0$ for $t<s_{\vee I}$ where
$s_{\vee I}:=\max s_I$ with $s_{\emptyset}=0$.

We suppose that the dynamics of $S^I$ is given by
\begin{equation}\label{SI}
dS_t^I(s_I)= S_t^I(s_I)*(\mu^I_t(s_I)dt+\Sigma_t^I(s_I)dW_t), \quad t> s_{\vee I}
\end{equation}
where $W$ is a $N$-dimensional  Brownian motion with respect to the filtration $\mathbb F$, the coefficients $\mu_t^I(s_I)$ and $\Sigma_t^I(s_I)$ are $\mathcal O_{\mathbb F}\otimes\mathcal B(\R_+^I) $-measurable and bounded. Note that for two vectors $x=(x_1,\cdots,x_N)$ and $y=(y_1,\cdots, y_N)$ in $\R^N$,
the expression $x*y$ denotes the vector  $(x_1y_1,\cdots,x_Ny_N)$.

We also suppose that one default event induces a jump on each remaining asset in the portfolio. More precisely, for any $I\neq\emptyset$, let
\begin{equation}\label{Equ:initial}
S^{I}_{s_{\vee I}}(s_I)=S_{s_{\vee I}-}^{J}(s_{J})*(\mathbf{1}-\gamma^{J,k}_{s_{\vee I}}(s_J)),
\end{equation}
where $k=\min\{i\in I|s_i=s_{\vee I}\}$, $J=I\setminus\{k\}$ and
$\gamma^{J,k}(\cdot)$ is $\mathcal P_{\mathbb F}\otimes\mathcal
B(\R_+^J)$-measurable, representing the jump  at $\tau_{\vee
I}=s_{\vee I}$ given the past default events $\tau_I=s_I$ with the
last arriving default $\tau_k$. The specification of $k$ and $J$
is for treating the case where several components in $s_I$ are
equal to $s_{\vee I}$. We remind that this convention is harmless
since we have assumed that $\tau_i\neq\tau_j$ a.s. if $i\neq j$.

The value of $S_T^I(\cdot)$ is determined by $S_{s_{\vee
I}}^I(\cdot)$ and the coefficients $\mu^I(\cdot)$ and
$\Sigma^I(\cdot)$. From a recursive point of view, $S_T^I(\cdot)$
actually depends on the initial value $S_0$ and the coefficients
indexed by all $J\subset I$, together with the jumps at defaults.

\begin{Rem}
We do not suppose that $\mathbb F$ is generated by the Brownian
motion $W$, allowing for some further generalizations.
Furthermore, we do not specify the set of assets and the set of
defaultable names, which permits to include a large family of
models.\end{Rem}

We give below several examples.

\begin{Exe} (Exogenous portfolio) We consider an exogenous investment portfolio, that is, the underlying assets in the investment portfolio are not included in the defaultable portfolio. In this case, the default family contains often highly risky names while the investors prefer to  choose assets in relatively better situations. However, these assets are influenced by the defaults. This is the case considered in \cite{JP2009}.
\end{Exe}

\begin{Exe}(Multilateral counterparty risks) The defaults family and the assets family coincide,  each underlying name subjected to the
default risk of itself and to the counterparty default risks of
the other names of the portfolio. For each name $i\in\Theta$,
denote by $S^i$ its value process and by $\tau_i$ its default
time. We suppose that the value of $S^i$ drops to zero at the
default time $\tau_i$, and at the default times $\tau_j$ where $
j\neq i, j\in\Theta$, the value of $S^i$ has a jump. So $S^i$ has
the decomposed form
$$S_t^i=\sum_{I \not\ni i}\indic_{A_t^I}S_t^{i,I}(\tau_I),$$
where $S_t^{i,I}(\cdot)$  is $\F_t\otimes\mathcal
B(\R_+^I)$-measurable. For $I\subset\Theta$ such that $i\notin I$,
let the dynamics of $S^{i,I}$ satisfy
\[\begin{split}
dS_t^{i,I}(s_I)=S_t^{i,I}(s_I)\big(\mu_t^{i,I}(s_I)dt
+\sigma_t^{i,I}(s_I)dW^i_t\big),\quad t> s_{\vee I}\end{split}\]
where $(W^1,\cdots, W^n)$ is an $n$-dimensional Brownian motion with covariance matrice $\Sigma$, $\mu_t^{i,I}(\cdot)$ and $\sigma_t^{i,I}(\cdot)$ are $\mathcal O_{\mathbb F}\otimes\mathcal B(R_+^I)$-measurable and bounded.
In addition, we suppose that for any $I\neq\emptyset$,
\[S^{i,I}_{s_{\vee I}}(s_I)=S_{s_{\vee I}-}^{i,J}(s_{J})(1-\gamma^{i,J,k}_{s_{\vee I}}(s_J))\]
where $k=\min\{i\in I|s_i=s_{\vee I}\}$, $J=I\setminus\{k\}$.

\end{Exe}

\subsection{A recursive optimization methodology}
In this subsection, we consider the optimal investment problem in
the contagion risk model. Following the recursive point of view on
the successive defaults in \cite{ejj2}, we have proposed in
\cite{JP2009} a two-steps --- before-default and after-default ---
optimization procedure. By using the density approach introduced
in \cite{ejj}, the optimizations are effectuated with respect to
the default-free filtration $\mathbb F$ instead of the global one
$\mathbb G$. This methodology has been adopted recently in
\cite{Pham2009} to add a  random mark at each default time by
using the joint density of the ordered default times and the
marks. We discuss this case as an application of constrained
optimization problems (Remark \ref{Rem fin}) using auxiliary
filtrations.

We apply now this recursive optimization methodology to the
contagion model described previously. We shall assume the density
hypothesis \ref{hyp:density} in the sequel with $\nu^{\btau}$, the law of $\btau$, being the Lebesgue measure. Namely, we assume that
the random measure of $\boldsymbol{\tau}$ is absolutely continuous
with respect to $\mathbb P\otimes d\bs$ on $\sigma$-field
$\mathcal F_t\otimes\mathcal B(\mathbb R_+^n)$, $t\in\R_+$, where
$d\bs$ is the probability law of $\boldsymbol{\tau}$. Denote by
$\alpha_t(\bs)$ the density of $\mu^{\btau}$ on $\mathcal
F_t\otimes\mathcal B(\mathbb R_+^n)$. By taking a suitable
version, one has that $\alpha(\bs)$ is an $\mathbb F$-martingale
for any $\bs\in\mathbb R_+^n$. Note that we do not suppose that
the defaults are ordered.

Let us consider an investor who holds a portfolio of assets, each
one subjected to contagion risks. The value process of the assets
is supposed to satisfy \eqref{SI} and \eqref{Equ:initial}. The
wealth process of this investor  is described by a positive
$\mathbb G$-adapted process $X$ and the allocation of the
portfolio is chosen by the criterion of maximizing the utility of
the terminal wealth at a finite horizon $T$. So we are interested
in the optimization problem $\esp[U(X_T)]$ where $U$ is a utility
function satisfying the conditions: strictly increasing, strictly
concave, $C^1$ on $\R_+$ and  the Inada conditions $U'(0_+)=\infty$, $U'(\infty)=0$.

The  portfolio of assets is characterized by a $\mathbb
G$-predictable process $\pi$, representing the proportion of the
wealth invested on each asset.  By Lemma \ref{Lem:decomopt}, the
process $\pi$ has the decomposed form
$\pi_t=\sum_I\indic_{A_{t-}^I}\pi_t^I(\tau_I)$ where
$\pi^I(\cdot)$ is $\mathcal P_{\mathbb F}\otimes\mathcal
B(\R_+^I)$-measurable, representing the investment strategy given
the defaults $\tau_I$. Hence, to determine an investment strategy
$\pi$ is equivalent to find a family
$(\pi^I(\cdot))_{I\subset\Theta}$.

Similarly,
the wealth process $X$ has the decomposed form $X_t=\sum \indic_{A_t^I}X_t^I(\tau_I)$ where $X^I(\cdot)$ is $\mathcal O_{\mathbb F}\otimes\mathcal B(\R_+^I)$-measurable, representing the wealth given $\tau_I$.  In view of \eqref{SI}, the wealth process dynamics is given by
\begin{equation}\label{XI}
 dX_t^I(s_I)=X_t^I(s_I)\pi_t^I(s_I)\cdot\big(\mu_t^I(s_I)dt+\Sigma_t^I(s_I)dW_t\big), \quad t>s_{\vee I}.
\end{equation} Note that for two vectors $x=(x_1,\cdots,x_N)$ and $y=(y_1,\cdots,y_N)$ in $\mathbb R^N$, the expression $x\cdot y$ denotes $x_1y_1+\cdots+x_Ny_N$.
By \eqref{Equ:initial}, we have for any $I\neq\emptyset$,
\begin{equation}\label{Equ:XIinit}X_{s_{\vee I}}^{I}(s_I)= X_{s_{\vee I}-}^J(s_J)(1-\pi_{s_{\vee I}}^{J}(s_J)\cdot\gamma_{s_{\vee I}}^{J,k}),\end{equation}
where $k=\min\{i\in I|s_i=s_{\vee I}\}$ and $J=I\setminus\{k\}$. To ensure that the wealth is positive, we need to suppose that the cumulative (proportional) losses caused by one default on all remaining names is smaller than $1$.

This following simple result is useful for the recursive optimization.
\begin{Lem} \label{Lem UXT}For any $T\in\R_+$,
\begin{equation}\label{Equ:calculU}\esp[U(X_T)]=\sum_{I\subset\Theta}
\int_{[0,T]^I\times]T,\infty[^{I^c}} \mathbb
E[U(X_T^I(s_I))\alpha_T(\boldsymbol{s})]d\boldsymbol{s}.\end{equation}
\end{Lem}
\begin{proof}We use the decomposed form of $X_T$ in Remark \ref{Rem2.8} and take iterated conditional expectation to obtain
$$\esp[U(X_T)]=\sum_{I\subset\Theta}\esp[\indic_{A_T^I}U(X_T^I(\tau_I))]=\sum_{I\subset\Theta} \esp\big[ \esp[\indic_{A_T^I}U(X_T^I(\tau_I))|\F_T] \big]. $$ The lemma then follows by definition of the density.
\end{proof}

We introduce the admissible strategy sets.
\begin{Def}\label{Def A}
For $I\subset\Theta$ and $s_I\in[0,T]^I$, let $\mathcal A^I(s_I)$
be the set of  $\mathbb F$-predictable processes $\pi^I(s_I)$ such
that the following two conditions are satisfied:
\begin{enumerate}[1)]
\item $\int_0^T|\pi^I_t(s_I)\sigma^I_t(s_I)|^2dt <\infty$;
\item in the case where $I\neq\Theta$, for any $i\notin I$ and any $s_i\in]s_{\vee I},T]$,
one has $\pi_{s_i}^{I}(s_I)\cdot\gamma_{s_i}^{I,i}<1.$
\end{enumerate}

Denote by $\mathcal A=\{(\pi^I(\cdot))_{I\subset\Theta}\}$ the set
of strategy families $\pi=(\pi^I(\cdot))_{I\subset\Theta}$, where
$\pi^I(\cdot)$ is a $\mathcal P_{\mathbb F}\otimes\mathcal
B(\R_+^I)$-measurable function such that for any $s_I\in[0,T]^I$,
the process $\pi^I(s_I)$ is in $\mathcal A^I(s_I)$. We say that
$\pi$ is admissible if
$\pi=(\pi^I(\cdot))_{I\subset\Theta}\in\mathcal A$.

\end{Def}

For our recursive methodology, it will also be useful to consider
all the strategies after the defaults $\tau_I=s_I$.   We introduce
the corresponding admissible sets below.

\begin{Def}\label{Def:A}
For any $I\subset\Theta$ and any $s_I\in[0,T]^I$, let $\mathcal
A^{\supset I}(s_I)$ be the set of families
$(\pi^K(s_I,\cdot))_{\,K\supset I}$, where $\pi^K(s_I,\cdot)$ is a
$\mathcal P_{\mathbb F}\otimes\mathcal B(\R_+^{K\setminus I}
)$-measurable function such that $\pi^K(s_K)\in\mathcal A^K(s_K)$
for any $s_{K\setminus I}\in[s_{\vee I}, T]^{K\setminus I}$.

We define the set $\mathcal A^{\supsetneq I}(s_I)$ in a similar
way. Note that any family $\pi^{\supset I}$ in $\mathcal
A^{\supset I}(s_I)$ can be written as $(\pi^I,\pi^{\supsetneq I})$
with $\pi^{\supsetneq I}\in\mathcal A^{\supsetneq I }(s_I)$ and
$\pi^I\in\mathcal A^{I}(s_I)$.
\end{Def}

Let us now  consider the maximization of the utility function on the terminal value of the wealth process
\begin{equation}\label{Equ: J0}
J(x,\pi):=\mathbb E[U(X_T)]_{X_0=x}.\end{equation}
We shall treat the optimization problem in a backward and recursive way. To this end,
we introduce some notations. Let
\[J_{\Theta}(x,\boldsymbol{s},\pi^{\Theta}):=
\mathbb E[U(X_T^{\Theta}(\boldsymbol{s}))\alpha_T(\bs)\,|\,
\mathcal F_{s_{\vee\Theta}}]_{X_{s_{\vee\Theta}}^{\Theta}(\bs)=x},\]
and
\[V_{\Theta}(x,\bs)=\underset{\pi^\Theta\in\mathcal A^\Theta(\bs)}\esssup\,J_{\Theta}(x,\boldsymbol{s},\pi^{\Theta}).\]
We define recursively for $I\subset\Theta$,
\begin{equation}\label{JI}\begin{split}J_I(x,s_I,\pi^I)&:=\mathbb E\bigg[
U(X_T^I(s_I))\int_{]T,+\infty[^{I^c}}\alpha_T(\bs)ds_{I^c}\\
&+
\sum_{i\in I^c}\int_{]s_{\vee I},T]}V_{I\cup\{i\}}
\big(X_{s_i}^{I\cup\{i\}}(s_{I\cup\{i\}}),s_{I\cup\{i\}}\big)ds_i\,\bigg|\,\mathcal F_{s_{\vee I}}\bigg]_{X_{s_{\vee I}}^{I}(s_I)=x},
\end{split}
\end{equation}
and correspondingly
\begin{equation}\label{VI}V_I(x,s_I):=\underset{\pi^I\in\mathcal
A^I(s_I)}{\esssup}\,J_I(x,s_I,\pi^{ I}).\end{equation}
\begin{Rem}\label{Rem:X}
Note that viewed from the initial time $t=0$, the value of
$X_T^I(s_I)$ depends on $X_0$ and all strategies $\pi^{J}$, $J\subset I$. 
However, viewed from the last arriving default
$\tau_{\vee I}$, its value depends only on the strategy $\pi^I$ if
the value of $X_{s_{\vee I}}^I(s_I)$ is given.
\end{Rem}

The above constructions provide us a family of optimization
problems $(V_I(x,s_I))_{I\subset\Theta}$.  Notice that at each
step, the optimization problem $V_I$ involves the resolution of
other ones $V_{I\cup\{i\}}$. The whole system need to be dealt
with in a recursive manner backwardly, each problem concerning the
filtration $\mathbb F$ and the time interval $[s_{\vee I}, T]$. By
resolving recursively the problems, we can obtain a family of
optimal strategies $(\widehat\pi^I(\cdot))_{I\subset\Theta}$.  The
following theorem shows that the global optimization problem,
which consists of finding the optimal strategy
$\widehat\pi\in\mathcal A$ for \eqref{Equ: J0} is equivalent to
finding $(\widehat\pi_I(\cdot))_{I\subset\Theta}$.

With the above notations, one has  in particular
$$J_{\emptyset}(x,\pi^\emptyset)= \esp\Big[U(X_T^\emptyset)\int_{]T,\infty[^n}\alpha_T(\bs)d\bs+\sum_{i=1}^n\int_{]0,T]}V_{\{i\}}(X_{s_i}^{\{i\}}(s_i),s_i)ds_i\Big]_{X_0=x}$$ and
$V_{\emptyset}(x)=\displaystyle\sup_{\pi^\emptyset\in\mathcal A^{\emptyset}}J_{\emptyset}(x,\pi^\emptyset)$.

\begin{Thm}\label{Thm opti}Suppose that  $V^I(x,s_I)<\infty$ a.s. for any $I\subset\Theta$, any $x>0$ and $s_I\in[0,T]^I$, then
\begin{equation}\label{Equ:them}\underset{\pi\in\mathcal A}{\sup}\, J(x,\pi)=V_{\emptyset}(x).\end{equation}
\end{Thm}
\begin{proof}For any $I\subset\Theta$, we deduce from a backward point of view and introduce
\begin{equation}\label{tilde J}\widetilde J_I(y,s_I,\pi^{\supset I})=
\mathbb E\Big[\sum_{K\supset I}
\int_{]T,\infty[^{K^c}}\int_{]s_{\vee I},T]^{K\setminus
I}}U(X_T^K(s_K))\alpha_T(\boldsymbol{s})ds_{I^c}\,\big|\,\mathcal
F_{s_{\vee I}}\Big],\end{equation} where $X_{s_{\vee I}}^I(s_I)=y$
and $\pi^{\supset I}$ is an element in $\mathcal A^{\supset
I}(s_I)$ (see Definition \ref{Def:A}). Note that the value of
$\widetilde J_I(y,s_I,\pi^{\supset I})$ depends on $y$, $s_I$ and
on the choice of $\pi^{K}(t_K)$ with $K\supset I$ and $t_I=s_I$.
We shall prove by induction  the equality
\begin{equation}\label{Equ:opt}
\underset{\pi^{\supset I}\in\mathcal A^{\supset I
}(s_I)}{\esssup}\,\widetilde J_I(y,s_I,\pi^{\supset
I})=V_I(y,s_I).
\end{equation}
By Lemma \ref{Lem UXT}, we have
$\widetilde J_{\emptyset}(y,\pi)=\esp[U(X_T)]_{X_0=y}$.
So the particular case of
\eqref{Equ:opt} when $I=\emptyset$ is just what need to be proved.

We proceed by induction on $I$ and begin by $I=\Theta$. Observe
that $\widetilde J_{\Theta}(y,\boldsymbol{s},\pi^{\supset\Theta})=
J_{\Theta}(y,\boldsymbol{s} ,\pi^{\Theta})$. Hence  the equality
\eqref{Equ:opt} holds true by definition when $I=\Theta$. Let $I$
be a proper subset of $\Theta$. Assume that we have proved
\eqref{Equ:opt} for all $K\supsetneq I$. We claim and show below
that
\begin{equation}\label{Equ:J}\begin{split}&\widetilde J_I(y,s_I,\pi^{\supset I})=
\esp\Big[ U(X_T^I(s_I))\int_{]T,\infty[^{I^c}}
\alpha_T(\boldsymbol{s})ds_{I^c}\\
&\quad +\sum_{i\in I^c}\int_{]s_{\vee I},T]}
\widetilde J_{I\cup
\{i\}}(X_{s_i}^{I\cup\{i\}}(s_{I\cup\{i\}}),s_{I\cup\{i\}},
\pi^{\supset I\cup\{i\}})ds_i|\mathcal F_{s_{\vee
I}}\Big]_{X_{s_{\vee I}}^I(s_I)=y}.\end{split}\end{equation}
In fact, by \eqref{tilde J}, the
second term in the right-hand side of \eqref{Equ:J} equals
\[\begin{split}&\quad\;\sum_{i\in I^c}\int_{]s_{\vee I},T]}ds_i
\mathbb E[\widetilde J_{I\cup \{i\}}(X_{s_i}^{I\cup\{i\}}(s_{I\cup\{i\}}),s_{I\cup\{i\}},
\pi^{\supset I\cup\{i\}})|\mathcal F_{s_{\vee I}}]\\
&=\sum_{i\in I^c}\sum_{K\supset I\cup\{i\}}
\int_{]T,\infty[^{K^c}}\int_{]s_{\vee I},T]^{\{i\}}}\int_{]s_{i},T]^{K\setminus(I\cup\{i\})}}
\mathbb E[U(X_T^K(s_K))\alpha_T(\boldsymbol{s})|\mathcal F_{s_{\vee I}}]ds_{I^c}\\
&=\sum_{K\supsetneq I}\sum_{i\in K\setminus
I}\int_{]T,\infty[^{K^c}}\int_{]s_{\vee
I},T]^{\{i\}}}\int_{]s_i,T]^{K\setminus(I\cup\{i\})}} \mathbb
E[U(X_T^K(s_K))\alpha_T(\boldsymbol{s})|\mathcal F_{s_{\vee
I}}]ds_{I^c}.
\end{split}\]
We consider, for any $i\in K\setminus I$, the set $]s_{\vee
I},T]^{\{i\}}\times ]s_{i},T]^{K\setminus(I\cup\{i\})}$. Note that
the subsets of $]s_{\vee I},T]^{K\setminus I}$ of the following
form
\[\Gamma_i:=\{s_{K\setminus I}\,|\,\forall\,j\in K\setminus(I\cup\{i\}),\;s_{\vee I}<s_i<s_j\le T\}.\]
are disjoint, in addition, the set
\[]s_{\vee I},T]^{K\setminus I}\big\backslash\textstyle\bigcup_{i\in K\setminus I}\Gamma_i\]
is negligible for the Lebesgue measure. Hence we obtain
\[\begin{split}&\quad\;\sum_{i\in I^c}\int_{]s_{\vee I},T]}ds_i
\mathbb E[\widetilde J_{I\cup
\{i\}}(X_{s_i}^{I\cup\{s\}}(s_{I\cup\{i\}}),s_{I\cup\{i\}},
\pi^{\supset I\cup\{i\}})|\mathcal F_{s_{\vee
I}}]\\&=\sum_{K\supsetneq I}\int_{]T,\infty[^{K^c}}\int_{]s_{\vee
I},T]^{K\setminus I}}\mathbb
E[U(X_T^K(s_K))\alpha_T(\boldsymbol{s})|\mathcal F_{s_{\vee
I}}]ds_{I^c},
\end{split}\]
and hence \eqref{Equ:J} is established.

By the induction hypothesis, one has
\begin{equation}\label{induction}\underset{\pi^{\supset
I\cup\{i\}}\in\mathcal A^{\supset
I\cup\{i\}}(s_{I\cup\{i\}})}{\esssup}\,\widetilde
J_{I\cup\{i\}}(y,s_{I\cup\{i\}},\pi^{\supset
I\cup\{i\}})=V_{I\cup\{i\}}(y,s_{I\cup\{i\}}).
\end{equation}
Hence we have by \eqref{Equ:J}
\[\begin{split}&\underset{\pi^{\supset I}
\in\mathcal A^{\supset I}(s_I)}{\esssup}\,\widetilde
J_I(y,s_I,\pi^{\supset I})\leq \int_{]T,\infty[^{I^c}}\mathbb E[
U(X_T^I(s_I))\alpha_T(\boldsymbol{s})|\mathcal F_{s_{\vee I}}]ds_{I^c}\\
&\quad+\sum_{i\in I^c}\int_{]s_{\vee I},T]} \mathbb
E[V_{I\cup\{i\}}(
X_{s_i}^{I\cup\{i\}}(s_{I\cup\{i\}}),s_{I\cup\{i\}})| \mathcal
F_{s_{\vee I}}]ds_i,
\end{split}
\]
which, together with the definitions \eqref{JI} and \eqref{VI},
implies
\[\underset{\pi^{\supset I}\in\mathcal A^{\supset I}(s_I)}{\esssup}\,\widetilde
J_I(y,s_I,\pi^{\supset I})\leq V_I(y,s_I).\]

We still suppose the induction hypothesis \eqref{induction} for
the converse. For $i\notin I$,
$s_{I\cup\{i\}}\in[0,T]^{I\cup\{i\}}$, $\varepsilon>0$ and
$z\in\mathbb R$, there exists a family $\pi^{\supset
I\cup\{i\}}_{\varepsilon,(z,i)}\in\mathcal A^{\supset
I\cup\{i\}}(s_{I\cup\{i\}})$ such that
\[\widetilde J_{I\cup\{i\}}\Big(z,s_{I\cup\{i\}},\pi^{\supset I\cup\{i\}}_{\varepsilon,(z,i)}\Big)\ge V_{I\cup\{i\}}(z,s_{I\cup\{i\}})-\varepsilon.\]
 We fix $s^I$, $X_{s_{\vee I}}^I(s_I)=y$ and $\pi^I\in\mathcal A^I(s_I)$. By a measurable selection result,
 we can choose a family $\pi^{\supsetneq
I}_\varepsilon\in\mathcal A^{\supsetneq I}(s_I)$ such that
\[\pi^{K}_\varepsilon(s_K)=\pi^K_{\varepsilon,(z_i,i)}(s_K),\]
where $i$, $z_i$ and $s_{K\setminus I}$ satisfy the following conditions:
\begin{enumerate}[(1)]
\item $i\in I^c$, $s_{\vee I}<s_i$ and $s_i<s_j$ for all $j\in K\setminus(I\cup\{i\})$,
\item $z_i=X_{s_i}^{I\cup\{i\}}(s_{I\cup\{i\}})$.
\end{enumerate}
Note that the value of $\widetilde
J_{I\cup\{i\}}(z_i,s_{I\cup\{i\}},\pi^{\supset I\cup\{i\}})$ only
depends on $z_i$, $s_{I\cup\{i\}}$ and the choice of $\pi^K(t_K)$
for those $t_K$ such that $t_{I\cup\{i\}}=s_{I\cup\{i\}}$. We
obtain that
\[\widetilde J_{I\cup\{i\}}(z_i,s_{I\cup\{i\}},\pi^{\supset I\cup\{i\}}_\varepsilon)
\ge V_{I\cup\{i\}}(z_i,s_{I\cup\{i\}})-\varepsilon.\]
This implies, by comparing \eqref{JI} and \eqref{Equ:J}, that
\[\widetilde J_I(y,s_I,(\pi^I,\pi_\varepsilon^{\supsetneq I}))
\ge J_I(y,s_I,\pi^I)-\varepsilon T|I^c|.\] By taking the essential
supremum over $\pi^I\in\mathcal A^I(s_I)$, one obtains
\[\underset{\pi^{\supset I}\in\mathcal A^{\supset I}(s_I)}{\esssup}\,\widetilde J_I(y,s_I,\pi^{\supset I})\geq V_I(y,s_I)-\varepsilon T|I^c|.\]
Since $\varepsilon$ is arbitrary, we get the inequality
\[\underset{\pi^{\supset I}\in\mathcal A^{\supset I}(s_I)}{\esssup}\,\widetilde J_I(y,s_I,\pi^{\supset I})\geq V_I(y,s_I).\]
We hence established the equality \eqref{Equ:opt}.
\end{proof}

The existence and the explicit resolution of the optimization
problems $(V_I(x,s_I))_{I\subset\Theta}$ 
will be discussed in detail in a companion paper.

\begin{Rem}\label{Rem fin}
It is often useful to consider strategies with constraints. In
this case, the  admissible set $ {\mathcal A}_{\circ}^I(s_I)$ for
$\pi^I(s_I)$, $s_I\in[0,T]^I$ is a subset of $\mathcal A^I(s_I)$
and the admissible trading strategy set ${\mathcal A}_{\circ}$ for
$\pi$ is defined  similarly~: ${\mathcal
A}_{\circ}=\{(\pi^I(\cdot))_{I\subset\Theta}\}$ such that for any
$s_I\in[0,T]^I$, $\pi^I(s_I)\in\mathcal A_{\circ}(s_I)$. We can also
define $\mathcal A_{\circ}^{\supset I }(s_I)$ and $\mathcal
A_{\circ}^{\supsetneq I }(s_I)$ in a similar way. Note that Theorem
\ref{Thm opti} still holds for the constrained strategy. More
precisely, let us introduce in a backward and recursive way
$J_I^{\circ}(x,s_I,\pi^I)$ similarly as in \eqref{JI} and let
$V_I^{\circ}(x,s_I)=\esssup_{\pi^I\in\mathcal
A_I^{\circ}}J_I^{\circ}(x,s_I,\pi^I )$. Then
$$\underset{\pi\in\mathcal A_{\circ}}{\sup}\, J(x,\pi)=V^{\circ}_{\emptyset}(x).$$
\end{Rem}

As an application, we consider the case where the 
losses at default times $\btau=(\tau_1,\cdots,\tau_n)$ are associated to some 
$\mathcal G$-measurable
random variables $\bL=(L_1,\cdots,L_n)$. This is the case studied in \cite{Pham2009} supposing $\btau$ is a family of ordered default times and using the joint density of $(\btau,\bL)$ with respect to the default-free filtration. 
We recall briefly this model and show that it can be considered as a constrained optimal problem  mentioned in the above remark.

\begin{Exe}\label{exe:pham} Let $\mathbb F^\circ$ be a default-free filtration. The default information contains the knowledge on default times $\tau_i$, $(i\in\Theta)$, together with an associated mark $L_i$ taking values in some Polish space $E$.  So the global market information is described by the filtration
\[\mathbb G^{\circ}=\mathbb F^{\circ}\vee\mathbb D^1\vee\cdots\vee
\mathbb D^n\vee\mathbb D^{L_1}\vee\cdots\vee\mathbb D^{L_n},\]
where for any $i\in\Theta$,  the filtration $\mathbb D^i$ is as in Section \ref{sec prel}, and the filtration $\mathbb D^{L_i}=(\D_t^{L_i})_{t\geq 0}$ is defined by $\D_t^{L_i}=\sigma(L_i\indic_{\{\tau_i\le
s \}},\,s\le t)$ made right-continuous.  Note that any $\mathbb
G^\circ$-optional (resp. predictable) process $Z$ can be written in the form
\[Z_t=\sum_{I\subset\Theta}\indic_{A_t^I}Z_t^I(\tau_I,L_I), 
 \quad(\text{resp. } \,Z_t=\sum_{I\subset\Theta}\indic_{A_{t-}^I}Z_t^I(\tau_I,L_I),)\]
where $Z^I(\cdot)$ is a $\mathcal O_{\mathbb F^\circ}\otimes\mathcal
B(\mathbb R_+^I\times\mathbb R^I)$ (resp. $\mathcal P_{\mathbb F^\circ}\otimes\mathcal
B(\mathbb R_+^I\times\mathbb R^I)$)-measurable function, $L_I=(L_i)_{i\in I}$. 
In particular, the control process $\pi$ can be written as $$\pi_t=\sum_{I\subset\Theta}\indic_{A_{t-}^I}\pi_t^I(\tau_I,L_I)$$ and  the wealth process $X$ as 
$X_t=\sum_{I\subset\Theta}\indic_{A_t^I}X_t^I(\tau_I,L_I)$.
\end{Exe}

We explain below how to interpret the above model as a constrained optimization problem. 
The point is to introduce suitable  auxiliary filtrations. Let $\mathbb
F:=\mathbb F^\circ\vee\sigma(L_1,\cdots,L_n)$. It is the initial
enlargement of $\mathbb F^\circ$ by including the family of marks $\bL$. 
Define also $\mathbb F^I:=\mathbb F^{\circ}\vee\sigma(L_i,i\in I)$. Observe that the control $\pi^I(\cdot,L_I)$  is actually $\mathcal P_{\mathbb F^I}\otimes\mathcal B(\R_+^I)$-measurable and is hence $\mathcal P_{\mathbb F}\otimes\mathcal B(\R_+^I)$-measurable. 
The wealth $X^I(\cdot,L_I)$ is $\mathcal O_{\mathbb F^I}\otimes\mathcal B(\R_+^I)$- and hence $\mathcal O_{\mathbb F}\otimes\mathcal B(\R_+^I)$-measurable.  We introduce the filtration 
$\mathbb G=\mathbb F\vee\mathbb
D^1\vee\cdots\mathbb D^n$, which is the progressive enlargement of the filtration $\mathbb F$ with respect to the default filtrations. Note that  $\mathbb G\supset\mathbb G^{\circ}$.

The Example \ref{exe:pham} can be considered as a constrained problem by using the auxiliary filtrations $\mathbb F$  and $\mathbb G$. Indeed, an admissible control process $\pi$ has now the decomposed form $\pi_t=\sum_{I\subset\Theta}\indic_{A_{t-}^I}\pi_t(\tau_I)$,
where $\pi^I(s_I)$ is $\mathcal P_{\mathbb F}\otimes\mathcal B(\R_+^I)$-measurable and  $\pi^I(s_I)\in\mathcal A^I(s_I)$, the admissible set $\mathcal A^I(s_I)$ being defined in Definition \ref{Def A}. Let us now make precise the constrained admissible strategy sets: let $\mathcal A_\circ^I(s_I)$ be the subset of $\mathcal A^I(s_I)$ such that $\pi^I(s_I)$ is $\mathcal P_{\mathbb F^I}\otimes \mathcal B(\R_+^I)$-measurable if $\pi^I(s_I)\in\mathcal A_\circ^I(s_I)$. By Remark \ref{Rem fin}, we can apply Theorem \ref{Thm opti} to solve the problem.
We finally remark that we only need the density hypothesis of $\btau$ with respect to the
filtration $\mathbb F$ but not necessarily the stronger one on the existence of the joint density of $(\btau,\bL)$ with respect to $\mathbb F^\circ$.

\end{document}